\documentclass[printer]{aa} 
\usepackage{graphicx}
\usepackage{txfonts}
\usepackage{natbib}

\begin{document}

\title{Searching for Jet Rotation in Class 0/I Sources observed with GEMINI/GNIRS}

\author{
Deirdre Coffey \inst{1}
\and Francesca Bacciotti \inst{2}
\and Antonio Chrysostomou \inst{3}
\and Brunella Nisini \inst{4}
\and Chris Davis \inst{5}
}

\institute{
The Dublin Institute for Advanced Studies, 31 Fitzwilliam Place, Dublin 2, Ireland. dac@cp.dias.ie
\and I.N.A.F. - Osservatorio Astrofisico di Arcetri, Largo E. Fermi 5, 50125 Firenze, Italy. fran@arcetri.astro.it
\and Centre for Astrophysics Research, University of Hertfordshire, Hatfield, HERTS AL10 9AB, U.K. 
Present address: Joint Astronomy Centre, 660 North A'ohoku Place, Hilo, HI 96720, USA. a.chrysostomou@jach.hawaii.edu
\and I.N.A.F. - Osservatorio Astronomico di Roma, Via Frascati 33, 00044 Monteporzio Catone, Italy. nisini@oa-roma.inaf.it
\and Joint Astronomy Centre, University Park, Hilo, Hawaii 96720, U.S.A. c.davis@jach.hawaii.edu
}

\date{Received 30 December 2009 / Accepted 24 November 2010}

\abstract
{In recent years, there has been a number of detections of gradients 
in the radial velocity profile across jets from young stars. 
The significance of these results is considerable. 
They may be interpreted as a signature of jet rotation about its symmetry axis, thereby representing the only existing observational indications 
supporting the theory that jets extract angular momentum 
from star-disk systems. However, the possibility that we are indeed observing 
jet rotation in pre-main sequence systems is undergoing active debate. }
{To test the validity of a rotation argument, we must extend the survey to a larger sample, 
including younger sources. }
{We present the latest results of a radial velocity analysis on jets from Class 0 and I sources, 
using high resolution data from the infrared spectrograph GNIRS on GEMINI South. 
We obtained infrared spectra protostellar jets HH\,34, HH\,111-H, HH\,212 NK1 and SK1. }
{The [Fe II] emission was unresolved in all cases and so Doppler shifts across the jet width could not be accessed. 
The H$_2$ emission was resolved in all cases except HH\,34. Doppler profiles across the molecular emission were obtained, and gradients in radial velocity of typically 3\,km\,s$^{-1}$ identified. }
{Agreement with previous studies implies they may be interpreted as jet rotation, leading to toroidal velocity and angular momentum flux estimates of 1.5\,km\,s$^{-1}$ and 1$\times$10$^{-5}$ M$_{\odot}\,$yr$^{-1}$\,AU\,km\,s$^{-1}$ respectively. 
However, caution is needed. For example, emission is asymmetric across the jets from HH\,212 suggesting a more complex interpretation is warranted. Furthermore, observations for HH\,212 and HH\,111-H are conducted far from the source implying external influences are more likely to confuse the intrinsic flow kinematics. 
These observations demonstrate the difficulty of conducting this study 
from the ground, and highlight the necessity for high angular resolution via adaptive optics or space-based facilities. }



\titlerunning{Searching for Class 0/1 jet rotation with GEMINI}
\authorrunning{Coffey et al. 2010}

\maketitle 

\section{Introduction}
\label{introduction}

One of the fundamental problems of star formation theory is finding an explanation 
of how the excess angular momentum may be extracted from the 
accreting gas and dust so that it may continue to travel 
inwards and eventually accumulate on the newly forming star. 
With the realisation that observed shocked gas was in fact 
the result of impacting high velocity jets and outflows ejected 
from such stars (\citealp{Bally07}; \citealp{Ray07}), came 
the proposal that such ejecta could be the underlying 
vehicle for angular momentum transport. The theoretical 
foundations have long been laid in which jets are ejected 
via magneto-centrifugal forces (\citealp{Pudritz07}; \citealp{Shang07}), 
but without observational verification. It became clear that high angular 
resolution observations are required to test the theory, 
since all such models describe acceleration and collimation on scales of tens of AU. 

A significant observational breakthrough has been made in recent 
years by our team. We have conducted a series of studies revealing 
for the first time indications that we can observe the jet rotating. 
This combined endeavour may prove to be the long-awaited observational 
backing for the magneto-centrifugal theory of star formation. 
The first study constituted observations of the outflow from Class 0 
source HH\,212, in three slit positions stepped parallel to the flow direction, 
and revealed a difference in radial velocity of 2\,km\,s$^{-1}$ across the flow 
in the H$_2$ 2.12$\mu$m near-infrared (near IR) line at 6$\arcsec$ (2\,500\,AU) 
from the star \citep{Davis00}. This was taken as the first observational 
hint of jet rotation, given the context of a concurrent study of the HH\,212 
disk which revealed a radial velocity gradient in the same sense \citep{Wiseman01}. 
Independantly, other studies examined jets from less embedded Class II sources 
at optical and near-ultraviolet (near UV) wavelengths, and harnessed the high 
resolution of the Hubble Space Telescope (HST) in order to observe jets closer 
to their ejection point. The DG\,Tau and RW\,Aur jets, examined with a similar 
parallel slit configuration, revealed radial velocity gradients of 
5-20\,km\,s$^{-1}$ within 100\,AU of the star (\citealp{Bacciotti02}; \citealp{Woitas05}). 
Furthermore, these gradients were sustained for $\sim$\,100\,AU along both lobes of the RW\,Aur bipolar jet. 
This sustainability does not favour alternative interpretations such as 
jet precession or asymmetric shocking. Such positive indications of 
jet rotation required a survey. We therefore examined several T\,Tauri systems, 
to study a series of 8 jet targets including two bipolar jets. 
This time the slit was placed perpendicular to the flow, which maximises 
the opportunity of detecting a rotation signature by observing across the full jet width. 
It also maximises survey efficiency and avoids difficulties introduced by uneven slit 
illumination \citep{Woitas05}. Analysis in the optical and near UV consistently 
produced systematic radial velocities differences of typically 15-25\,km\,s$^{-1}$ 
close to the ejection point (\citealp{Coffey04}; 2007). These very encouraging 
statistics led to a pilot study to detect gradients in younger Class I sources, 
since rotation signatures should be present at all evolutionary stages. 
Indeed, near IR spectra of HH\,26 and HH\,72 revealed gradients in their radial 
velocity profiles transverse to the flow direction at 1\,000\,AU from the star \citep{Chrysostomou08}. 

In response to the positive outcome of this pilot study, 
we now conduct a wider survey of younger sources 
to consolidate the results. 
Three Class 0/I systems were observed in near IR lines, using the high spectral resolution (R$\sim$18\,000) available with GNIRS on GEMINI South. Based on previous detections of outflows in the near IR, we choose HH\,34 (\citealp{{Davis01}}; \citealp{{Davis03}}), 
HH\,111-H \citep{Davis01b}, and HH\,212 NK1 and SK1 \citep{Davis00}. 
The [\ion{Fe}{II}]\,1.64\,$\mu$m emission traces the hot inner parts of the jet 
while the H$_2$ emission at 2.12\,$\mu$m traces the warmer outer regions. 
Analysis of the spatially broad emission, which is resolved under good seeing conditions, 
takes full advantage of the high spectral resolution setting. 
Hence, using spectral analysis techniques, we are able to determine 
whether or not a Doppler gradient is present and in what sense, 
and thus the magnitude of the implied toroidal velocity. 

\section{Sample}
\label{sample}

\subsection{HH\,34}
\label{hh34}
The HH\,34 jet, first reported by \cite{Reipurth86}, is the name given to the most recent ejection of the highly collimated S-shaped parsec-scale HH\,34 flow from a Class\,I source in Orion, which is located at 414\,pc distant \citep{Menten07}. It has since been well studied via imaging and spectroscopy, but perhaps most impressive are the deep high resolution {\em HST} optical images \citep{Reipurth02}, used to determine morphology, photometry and excitation of the flow, and reveal that the jet abruptly changes direction possibly due to the powerful tidal effects of a companion star. This is in addition to the large scale S-shape of the flow already reported by \cite{Bally94} and \cite{Devine97}. Periodic time-variable ejection modeling of the jet have successfully reproduced the flow structure and morphology (\citealp{Raga98}; 2002c), while \cite{Masciadri02} investigate precession of the flow as a deceleration mechanism. More recently, high resolution optical integral field spectroscopy observations \citep{Beck07} show agreement with the kinematics and electron density structure predicted by existing internal working surface models, although radial velocity studies show no evidence of a Doppler gradient across the flow but this may be because the effective resolution of 20\,km\,s$^{-1}$ did not allow a detection. Close to the source, the H$_2$ counter-part to the optical jet emission was first traced by \citet{Davis01}, while the same region was also seen in [\ion{Fe}{II}] 1.64$\mu$m emission \citep{Davis03}. Further recent near-IR kinematic and diagnostic studies of the jet physics include \cite{Podio06}; \cite{Takami06}; \cite{GarciaLopez08}; \cite{Antoniucci08}. 

\subsection{HH\,111 - H}
\label{hh111}

The highly collimated HH\,111 Class I outflow was first reported by \cite{Reipurth89}. 
Located in Orion, the HH\,111 flow has been studied in detail over the years. 
Observations of the outflow itself at high resolution include HST optical images (\citealp{Hartigan01}; 
\citealp{Reipurth97}) and spectra \citep{Raga02}. As with HH\,34, it shows a chain of well aligned knots ending in a bowshock, the H knot being one of the brightest and located at 33$\arcsec$ (13700\,AU) from the source. Originating from a known multiple system, it is remarkable for its stability over parsec scales. It is also a strong H$_2$ emitter \citep{Gredel93}, and is associated with a powerful CO outflow \citep{Cernicharo96}. The H$_2$ flow has been examined using echelle spectroscopy \citep{Davis01b} which revealed knot H to have a double-peaked profile, interpreted with a simple, geometrical bow shock model. A sister bipolar flow, HH\,121 (\citealp{Gredel93}; 1994), originates from the same position but is offset in P.A. by 60\degr~from HH\,111. The core of this quadrupolar outflow has also been examined \citep{Rodriguez08} with a detection possibly indicating two disks. Models of the ejection history of HH\,111 western lobe (\citealp{Raga02}; \citealp{Masciadri02b}) show the bow shocks to be the result of an ejection velocity time-variability, while near IR spectral diagnostic studies \citep{Nisini02} demonstrate the nature of the shocked gas. 

\subsection{HH\,212 NK1 \& SK1}
\label{hh212}

HH\,212 is a well-known H$_2$ outflow from a Class 0 source in Orion. 
Early imaging observations \citep{Zinnecker98} clearly show a highly symmetric bipolar flow, 
with a total length of 240\arcsec (0.5\,pc). 
The H$_2$ knots closest to the source in each lobe ($\sim$6$\arcsec$) are named NK1 and SK1. 
The HH\,212 system is also associated with a collimated CO outflow \citep{Lee06}, and an SiO flow (\citealp{Codella07}; \citealp{Cabrit07}; \citealp{Lee07}; 2008). Near IR diagnostic studies include \citet{CarattioGaratti06}, which examines physical properties and cooling mechanisms, and \citet{Smith07}, which reveals excitation properties consistent with outward-moving bow shocks close to the plane of the sky. 

The first hint that jet rotation is detectable in a young stellar jet actually came with the observations of HH\,212 by \cite{Davis00}, based on a difference in the radial velocity across the receding lobe of the H$_2$ flow. This was interpreted as such given the agreement with the sense of a radial velocity gradient across the disk of the same system \citep{Wiseman01}. \cite{Codella07} report no sign of jet rotation in the SiO emission near NK1, whereas in the region SK2-SK4 ($\sim$10-14$\arcsec$ from source), they find a gradient in a direction contrary to that reported for SK1 by \cite{Davis00}. It is therefore also contrary to that of the disk which has a 5\,km\,s$^{-1}$ difference between the blue-shifted ammonia gas in the north-east and the red-shifted gas in the south-west \citep{Wiseman01}. However, \cite{Lee08} recently conducted a higher resolution rotation study of the HH\,212 flow and report matching gradients in the southern and northern lobes of the SiO jet in a sense that also matches that of the disk. They note that the slope of the gradient in the northern lobe (at 800\,AU from the disk plane) is smaller than the southern lobe (at 450\,AU). Indeed, a very comprehensive H$_2$ study of this system was conducted using GEMINI/Phoenix \citep{Correia09}, in which the various kinematic possibilities are modeled with the conclusion that the observations cannot be reproduced by jet rotation alone, though jet rotation may be included with other effects such as precession. Furthermore, near IR diagnostic studies \citep{Smith07} report a gradient in excitation transverse to the jet axis across in the inner knots, suggesting a transverse source motion rather than precession, possibly related to the jet bending and the transverse gradient in radial velocity. A similar gradient in excitation transverse to the jet axis in the T\,Tauri star Th~28 was reported by  \citet{Coffey08}, which possibly suggests asymmetric shocking. 

\begin{table*}
\begin{center}
\scriptsize{\begin{tabular}{llllcccccc}
\hline\hline
Target		&RA		&Dec		&PA$_{flow}$ 	&$v_{lsr}$	&$i_{jet}$	&References \\
		&(2000)		&(2000)		&(deg)		&(km\,s$^{-1}$)	&(deg)		&	\\ 
\hline
HH\,34		&05 35 29.8	&-06 26 58.0	&166 		&8.7 		&23-28	 	&1, 2, 3, 4 \\
HH\,111 H	&05 51 44.2	&+02 48 34.0	&97 		&23.0 		&10 		&5, 6  \\
HH\,212 NK1	&05 43 51.547	&-01 02 48.0  	&23		&1.7		&4		&7, 8 \\ 
HH\,212	SK1	&05 43 51.246	&-01 02 58.35 	&23	 	&1.7		&4		&7, 8 \\ 
\hline
\end{tabular}}
\end{center}
\caption{Targets investigated in this paper, 
all located in Orion at 414\,pc \citep{Menten07}. 
Our slit is placed perpendicular to the position angle (east of north) of the jet/outflow, PA$_{flow}$. The systemic velocity is taken as the cloud LSR velocity, $v_{lsr}$.  
The angle of inclination of the jet, $i_{jet}$, is given with respect to the plane of the sky. 
References - 
(1) \citealp{Reipurth02}; (2) \citealp{Anglada95}; (3) \citealp{Eisloffel92}; (4) \citealp{Heathcote92}; 
(5) \citealp{Reipurth89}; (6) \citealp{Reipurth92}; 
(7)  \citealp{Claussen98}; (8) \citealp{Wiseman01}. 
\label{targets}}
\end{table*}
\section{Observations}
\label{observations}

Observations were conducted, in queue mode, with the GEMINI Near InfraRed Spectrograph (GNIRS) on GEMINI South in mid-October 2006, late December 2006 and early January 2007. Near IR spectra were obtained of four jets from Class 0 and I sources (see Section\,\ref{sample}) for one slit position at a given distance along the outflow, and orientated perpendicular to the direction of propagation. Observations were made (with position angle measured east of north) of the HH\,34 jet at 1$\arcsec$ (400\,AU) from the source (P.A.$_{slit}$=77, 257\degr according to guide star availability on the observation date), and the H knot in the HH\,111 jet located at 34$\arcsec$ (14000\,AU) from the source (P.A.$_{slit}$=7\degr). Observations were also made of one knot in the approaching and receding jet from HH\,212, namely NK1 and SK1 respectively, both located at 6$\arcsec$ (2500\,AU) from the source (P.A.$_{slit}$=116\degr).

Using the long slit (49$\arcsec$) and the long-blue camera configuration, in combination with the 110.5 l/mm grating and choosing a 2-pixel slit-width (0$\farcs$10), we achieve a spectral resolving power of R\,$\sim$\,17\,800. The corresponding velocity resolution is 17\,km\,s$^{-1}$. Average seeing during the observations varied between 0$\farcs$5 and 0$\farcs$8. 

We can use profile fitting to achieve an effective spectral resolution which is beyond the actual resolution of the observations. 
Given a line profile which is intrinsically Gaussian, fitting allows an effective resolution which improves with climbing signal-to-noise. The one sigma error on the centroid is given by : 
\begin{eqnarray}
\sigma_{centroid} & = & \frac{resolution}{2.3548~snr}
\label{centroid}
\end{eqnarray}
where $resolution$ is either spatial or spectral, depending on the application, and $snr$ is the signal-to-noise ratio \citep{Whelan08}. 

We chose filters G0504 and G0503 respectively, in order to detect the emission of [\ion{Fe}{II}]\,1.64\,$\mu$m and H$_2$\,2.12\,$\mu$m. 
The standard procedure for observations in the near IR was adopted. Eight integrations of 300\,s were made, using the ABBA technique, and then co-added. Two such exposures in each filter were obtained, in order to reach sufficient signal-to-noise in the jet borders. The data were calibrated and reduced according to standard procedures, using the tools within IRAF specifically designed for the reduction of GEMINI data. This included running tasks to correct for non-linearity (\texttt{nsprepare}) and spatial distortions (\texttt{nsdist}). Wavelength calibration was achieved using a dispersion scale derived from arc lines (\texttt{nswavelength}), which straightened and dispersion corrected the exposure (\texttt{nstransform}). All velocities are adjusted to the LSR reference frame (\texttt{rvcorrect}) and further adjustment was made for the LSR velocity of each system, V$_{LSR}$, Table\,\ref{targets}. Additional overheads of standard star observations were not required for a radial velocity analysis, and so data are not flux calibrated. 

\section{Data Analysis}
\label{dataanalysis}

If protostellar jets are accelerated magneto-centrifugally then, with sufficient observing resolution and good signal-to-noise, it should be possible to detect the rotation signature of the flow. For good signal-to-noise, we can expect a typical accuracy in radial velocity of $\sim$1.5\,km\,s$^{-1}$. 
In the case of a rotating jet, the gas on one side of the flow axis is expected to display a higher Doppler shift than on the other side. 
Therefore, measuring a gradient in the radial velocity profile  
in the direction perpendicular to that of jet propagation 
can be interpreted as a signature of jet rotation. 
In such a case, the magnitude of the radial velocity differences (typically accurate to $\sim$2\,km\,s$^{-1}$) between the two sides of the flow axis allows derivation of a toroidal velocity component. 
Observationally, this requires that the jet axis is identifiable, 
and that high spatial and spectral velocity resolution is achieved 
to reveal the velocity profile as a function of distance from the jet axis. 
Clearly, this study is very demanding, 
and so obtaining useful observations from the ground is difficult. If the signal-to-noise is sufficiently high, however, we can reach beyond the nominal resolution via profile fitting of jet emission lines. 

As expected, the H$_2$ emission was found to be spatially broad, and so is almost always spatially resolved (with the exception of HH\,34). Therefore, we could profile fit in the dispersion direction to obtain a jet radial velocity profile transverse to the flow propagation, and thus determine the magnitude and direction of any implied jet toroidal velocity. First, the emission was binned and fitted with a Gaussian in the spatial direction, the peak of which was assumed to denote the jet axis. This was used as a reference point in determining differences in radial velocity between one side of the jet and the other.  Each pixel row of the unbinned emission was then fitted with a Gaussian in the spectral direction, and the measured radial velocities at mirrored positions either side of this central position were subtracted. In a couple of cases where the emission profile was not entirely symmetric, cross-correlation was deemed more appropriate. In this way, we find how any differences in Doppler shift vary with distance from the jet axis. 

For the [Fe\,II] emission, the spectral profile clearly traces both a high and low velocity component of jet material. 
Spatially, however, the emission is more confined to the jet axis and the seeing is never sufficiently clear for emission to be resolved implying the atomic jet width is at least $<$0$\farcs$6. Thus, in our seeing limited data, we cannot access the Doppler profile across the atomic component of the flow. 

\section{Results}
\label{results}

\subsection{HH\,34}
\begin{figure*}
\centering 
\includegraphics[width=6.5cm]{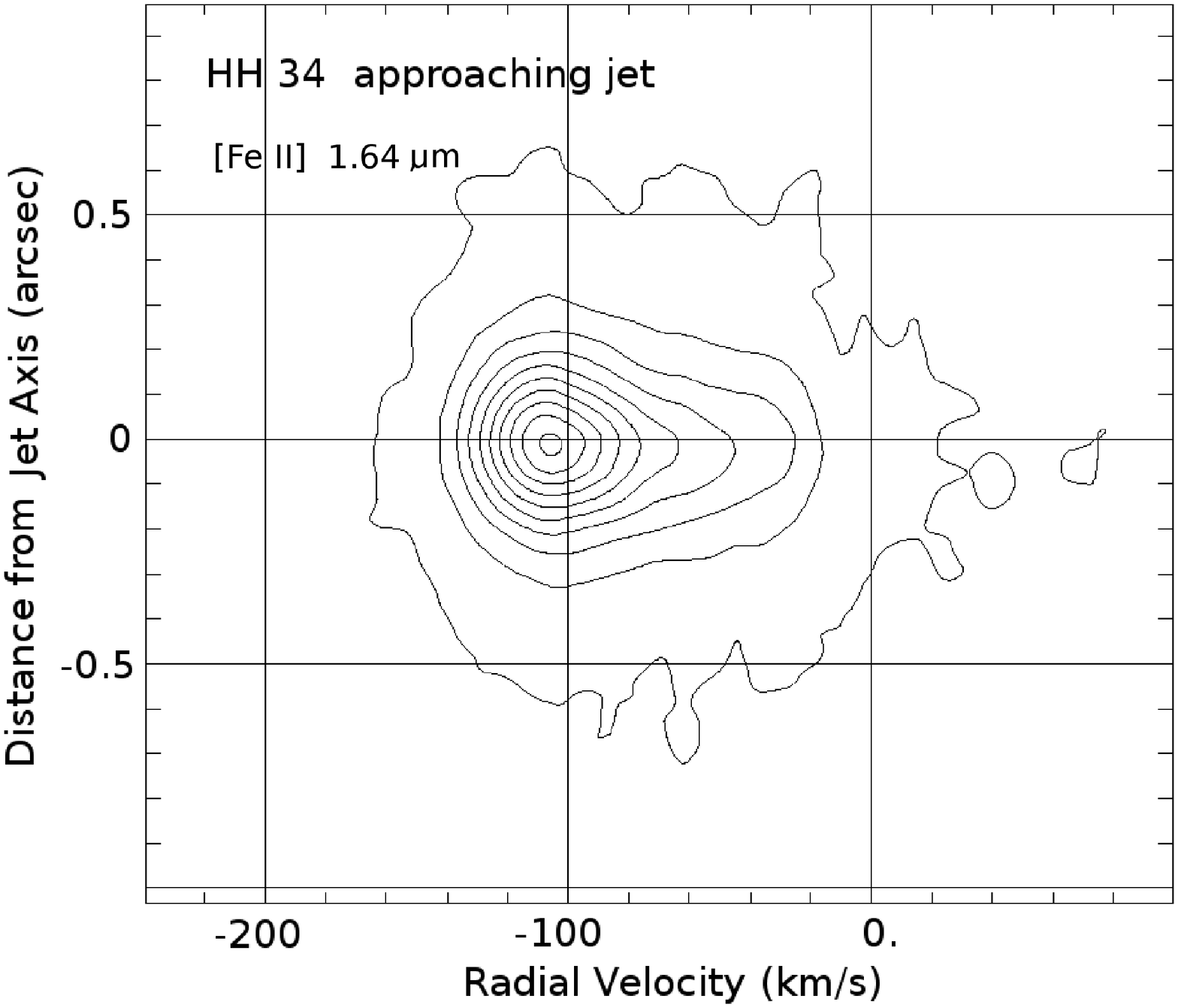}
\includegraphics[width=6.5cm]{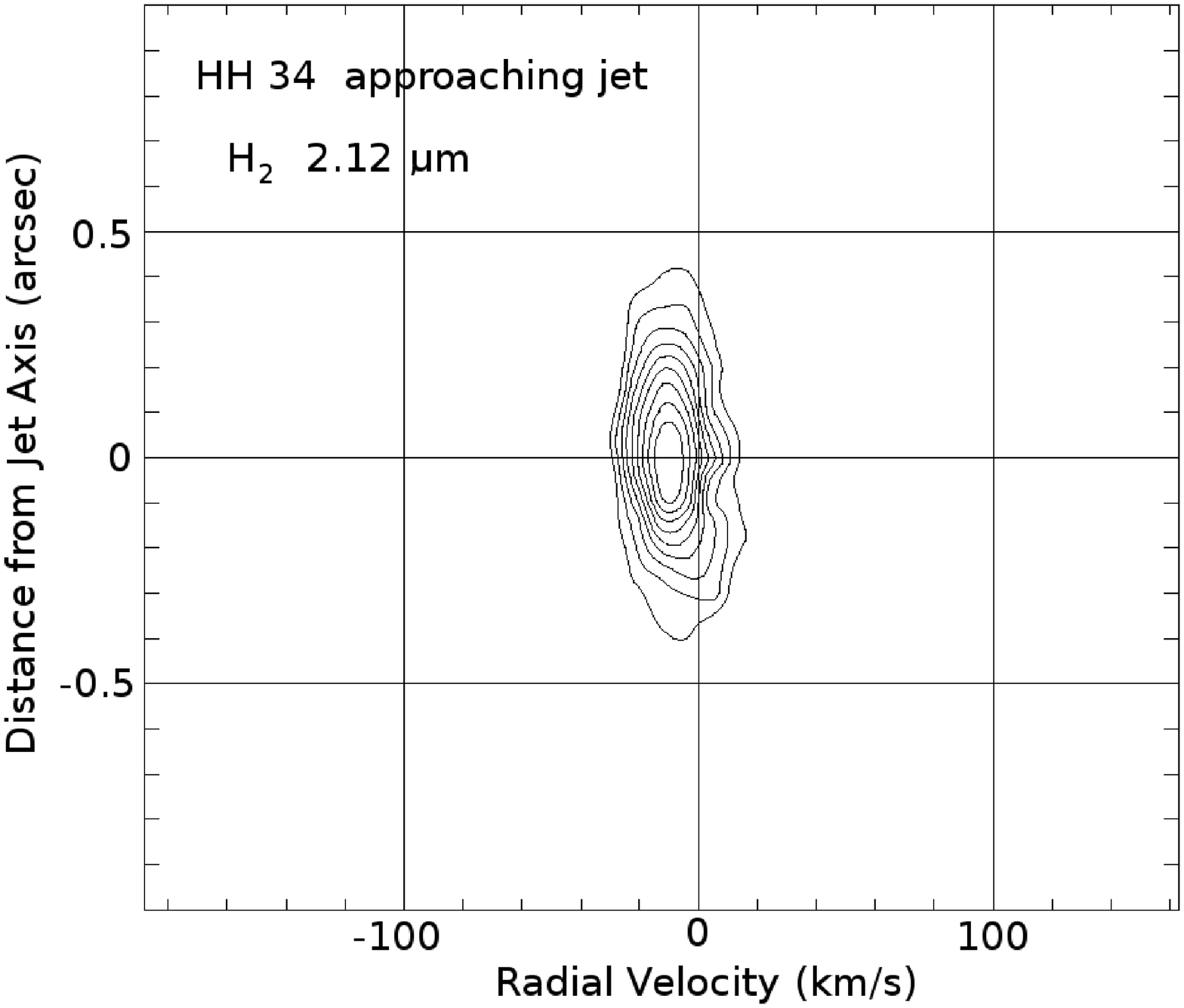}
\caption{Position-velocity diagram of [\ion{Fe}{II}] and H$_2$ emission from the HH\,34 approaching jet. Positive distances correspond to south-west. Both are spatially unresolved under the seeing of 0$\farcs$4 and 0$\farcs$6 respectively. 
\label{hh34_fe1}}
\end{figure*}

The HH\,34 jet was clearly detected in both [\ion{Fe}{II}] and H$_2$ emission. 
The observations are taken at only 1$\arcsec$ (400\,AU) above the disk-plane. 
Exposures for a given filter were {\em not} co-added, in order to preserve the information obtained in higher seeing conditions. 
Figure\,\ref{hh34_fe1} shows the position-velocity plots for one exposure, observed in October. The data contained reflected continuum emission with which the seeing at the time of observation could be measured. 
Unfortunately, the jet width is unresolved in both lines under the seeing of 0$\farcs$4 and 0$\farcs$6 respectively. 
The [\ion{Fe}{II}] higher velocity component (HVC) is travelling at -96\,($\pm$2)\,km\,s$^{-1}$, consistent with measurements of \citet{Davis03}; \citet{Takami06} and \citet{GarciaLopez08}. 
Meanwhile, the H$_2$ lower velocity component (LVC) is travelling at -15\,($\pm$1)\,km\,s$^{-1}$. 
The H$_2$ exposures show a deceleration from -15 to -10\,km\,s$^{-1}$ between October and December observations. 
There is currently no available measure on the sense of disk rotation for this system. 

\subsection{HH\,111 - H}
\begin{figure*}
\centering 
\includegraphics[width=6.5cm]{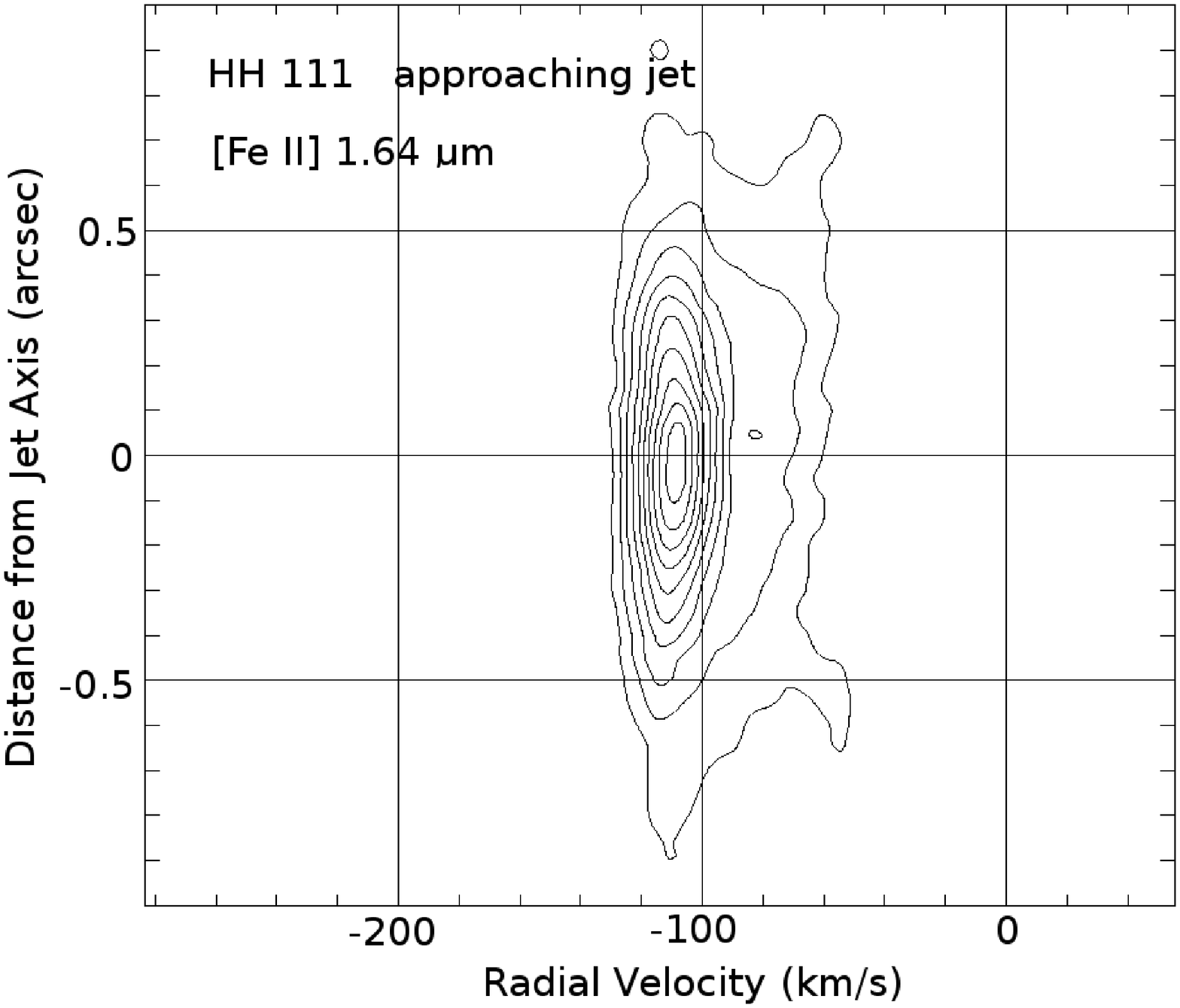}
\includegraphics[width=6.5cm]{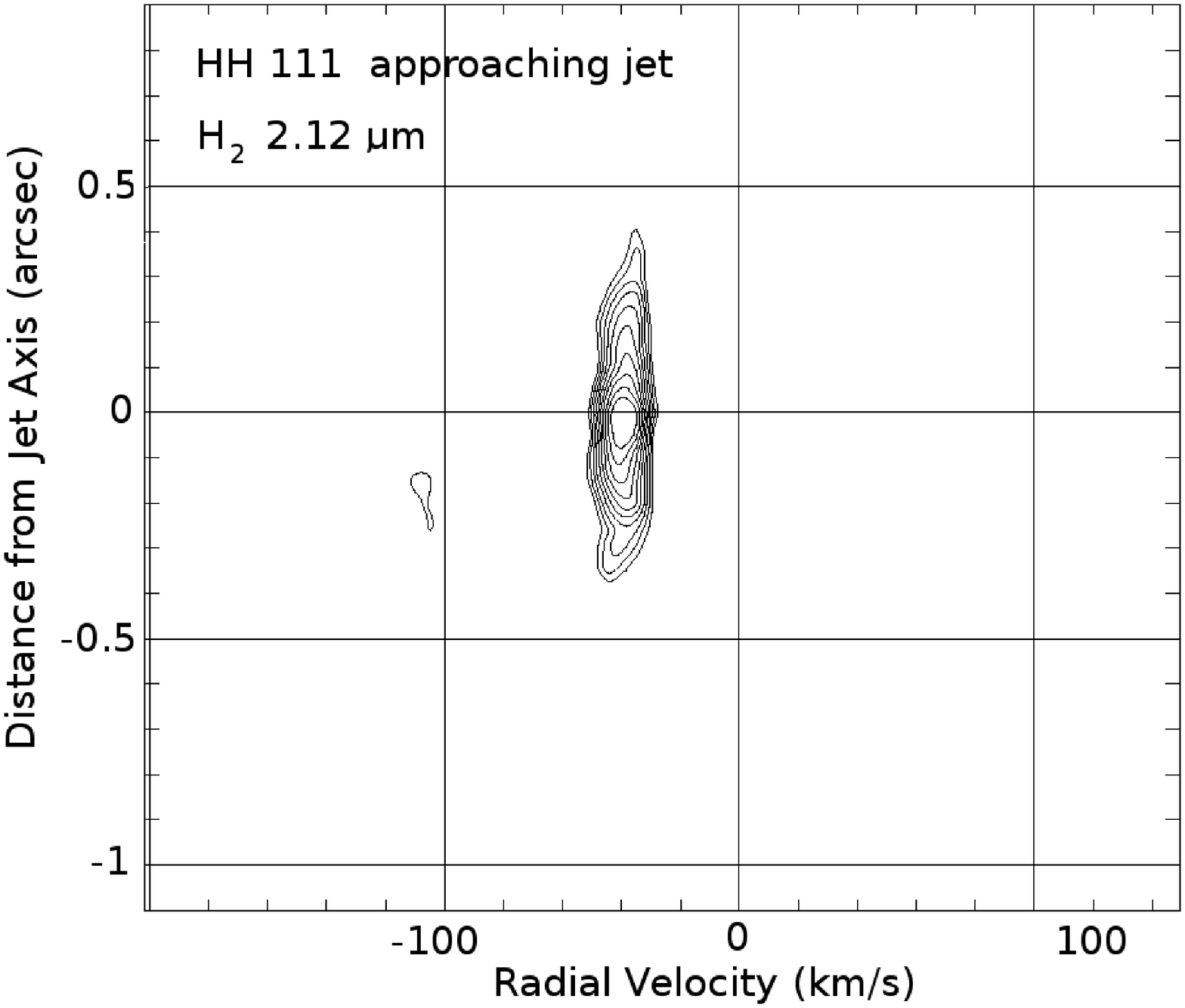}
\includegraphics[width=6.5cm]{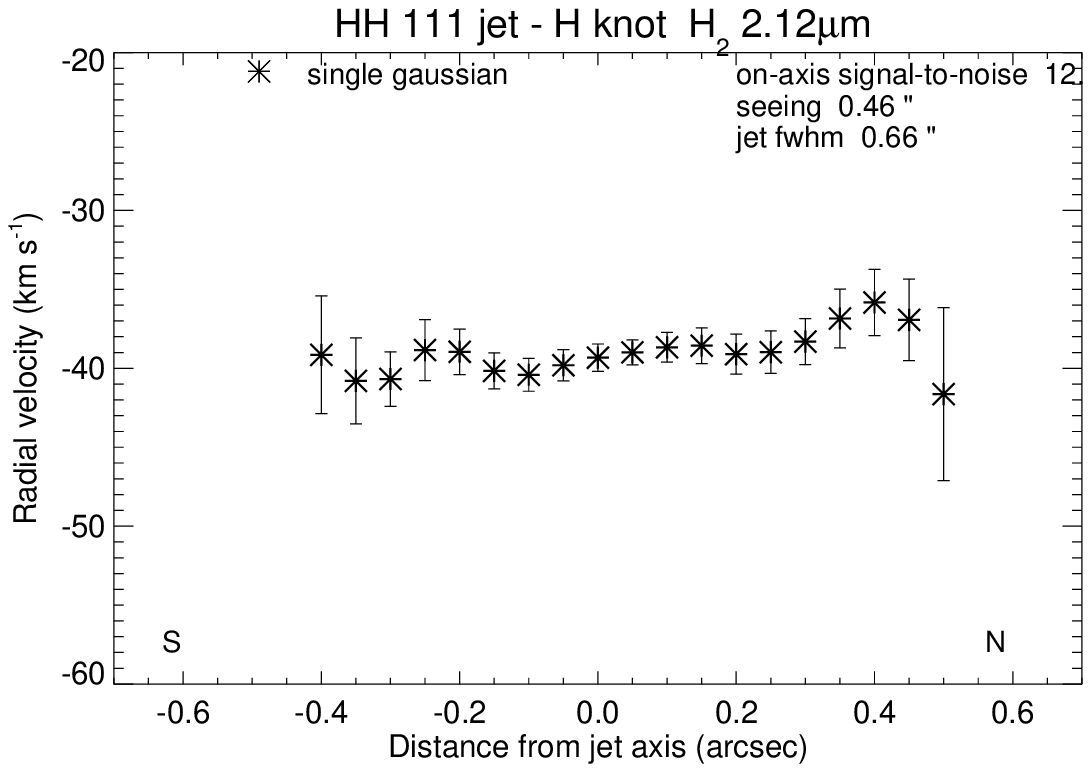} 
\includegraphics[width=6.5cm]{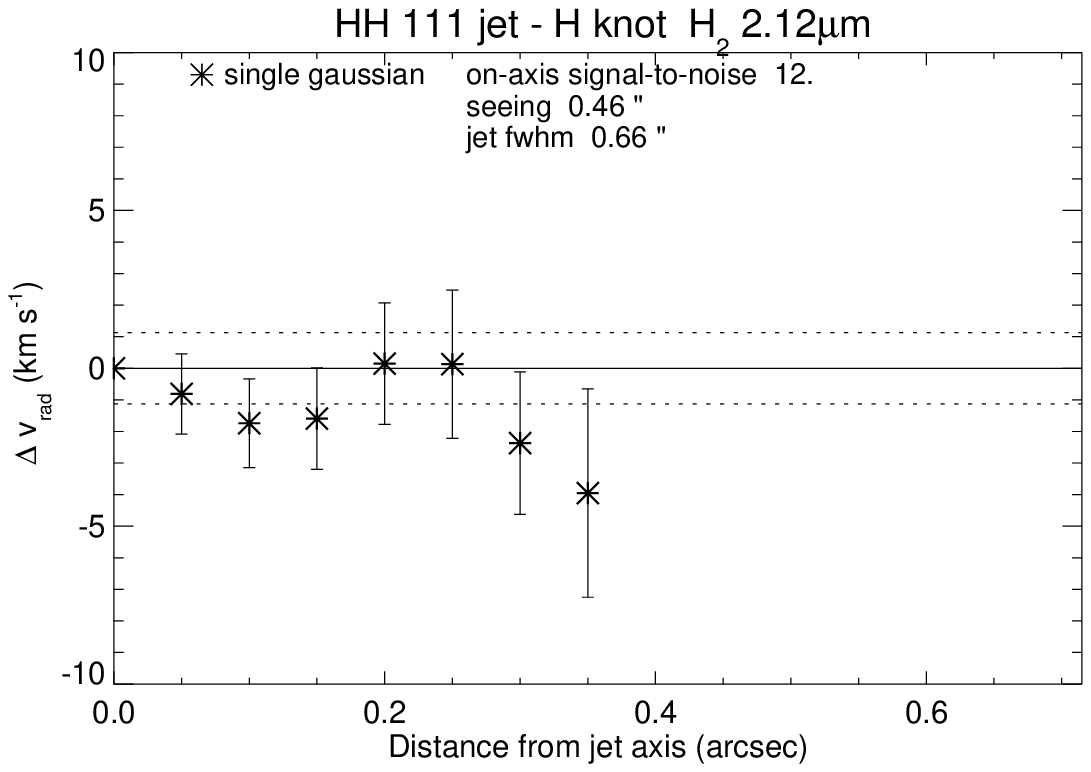}
\caption{Top: position-velocity plots for the H knot of the HH\,111 jet in [\ion{Fe}{II}] and H$_2$ emission, where positive distances correspond roughly to north; Bottom: radial velocity profile across the jet  in the spatially resolved H$_2$ emission, and a plot of the differences in the Doppler shift, $\Delta$v$_{rad}$, between one side of the jet knot and the other.
\label{hh111_fe}}
\end{figure*}

While the HH\,111 jet is clearly detected in both [\ion{Fe}{II}] and H$_2$ emission, Figure\,\ref{hh111_fe}, the signal-to-noise in H$_2$ was extremely poor in spite of the long integration. There was no reflected continuum emission in either line 
from which to determine the seeing during observations, and so we rely on the weather report at the time. 
The [\ion{Fe}{II}] HVC is travelling at -107\,($\pm$1)\,km\,s$^{-1}$, while the H$_2$ LVC is travelling at -39\,($\pm$2)\,km\,s$^{-1}$, in line with \citet{Davis01b}. 
Disk observations show that the northern side is blue-shifted with respect to the southern side (\citealp{Yang97}; \citealp{Lee09}). We therefore could expect to detect positive values of radial velocity differences, $\Delta$v$_{rad}$. Our results are all within error bars about zero, and so we do not detect a clear Doppler gradient. A mere hint of a gradient may be apparent, but these are in the opposite direction  (i.e. negative values of differences in radial velocity, $\Delta v_{rad}$) to that expected based on the sense of disk rotation measurements. Note that we are examining this jet knot at 33$\arcsec$ from the source and so environmental influences are more likely come into play in significantly disrupting the flow kinematics. 
Finally, in the position-velocity diagram, we see that a second emission peak in H$_2$ was observed (offset spatially by 0$\arcsec$2). The emission is much fainter and appears to trace an offset HVC in H$_2$, with V$_{LSR}$ measured as -110\,($\pm$3)\,km\,s$^{-1}$. A similar faint HVC is sometimes seen in echelle spectra of bow shocks, see e.g. \cite{Smith03} (on HH 7). In this case, the HVC is attributed to a fast-moving mach disk. 

\subsection{HH\,212 - NK1}
\label{hh212nk1}

\begin{figure*}
\centering 
\includegraphics[width=6.5cm]{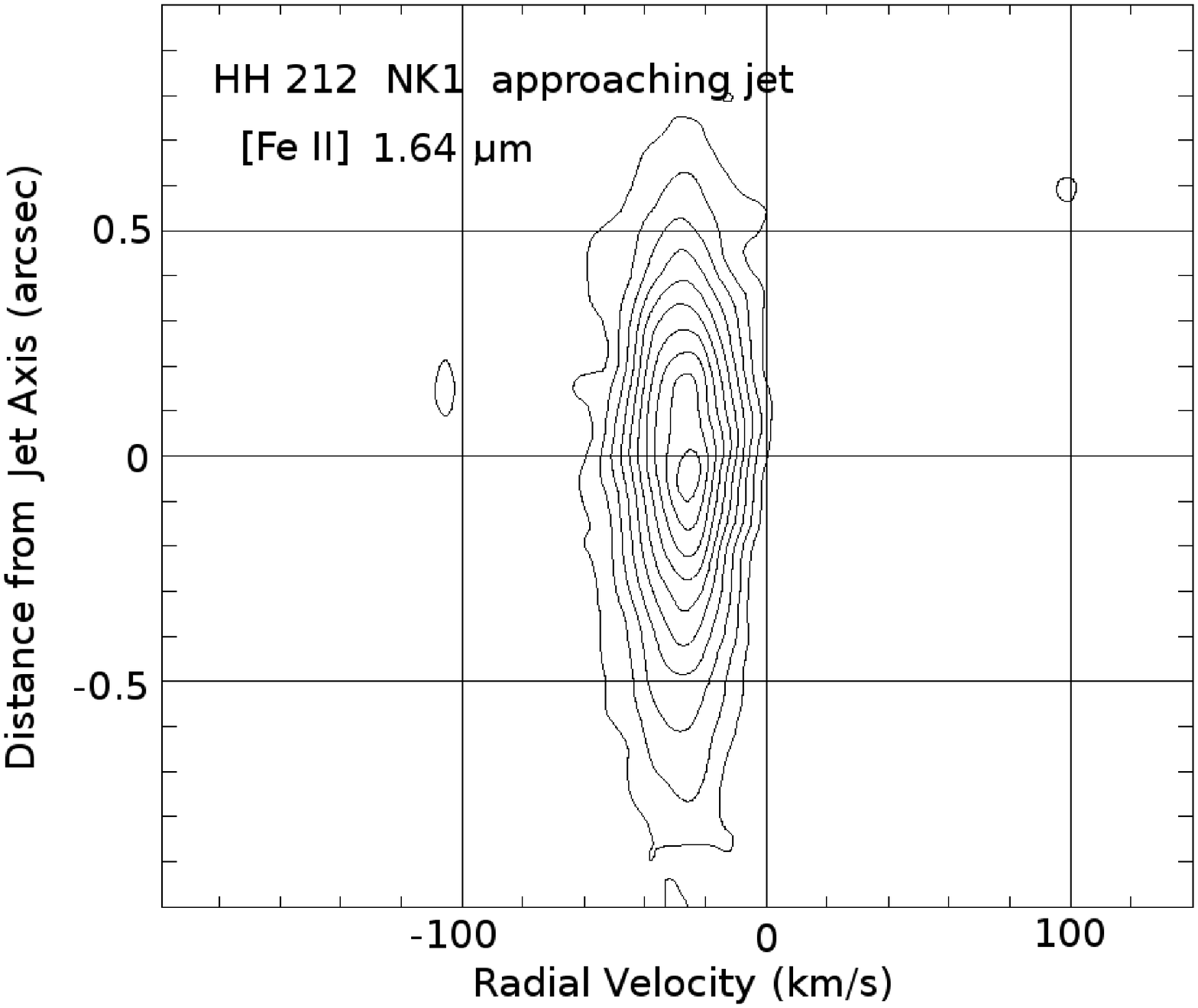}
\includegraphics[width=6.5cm]{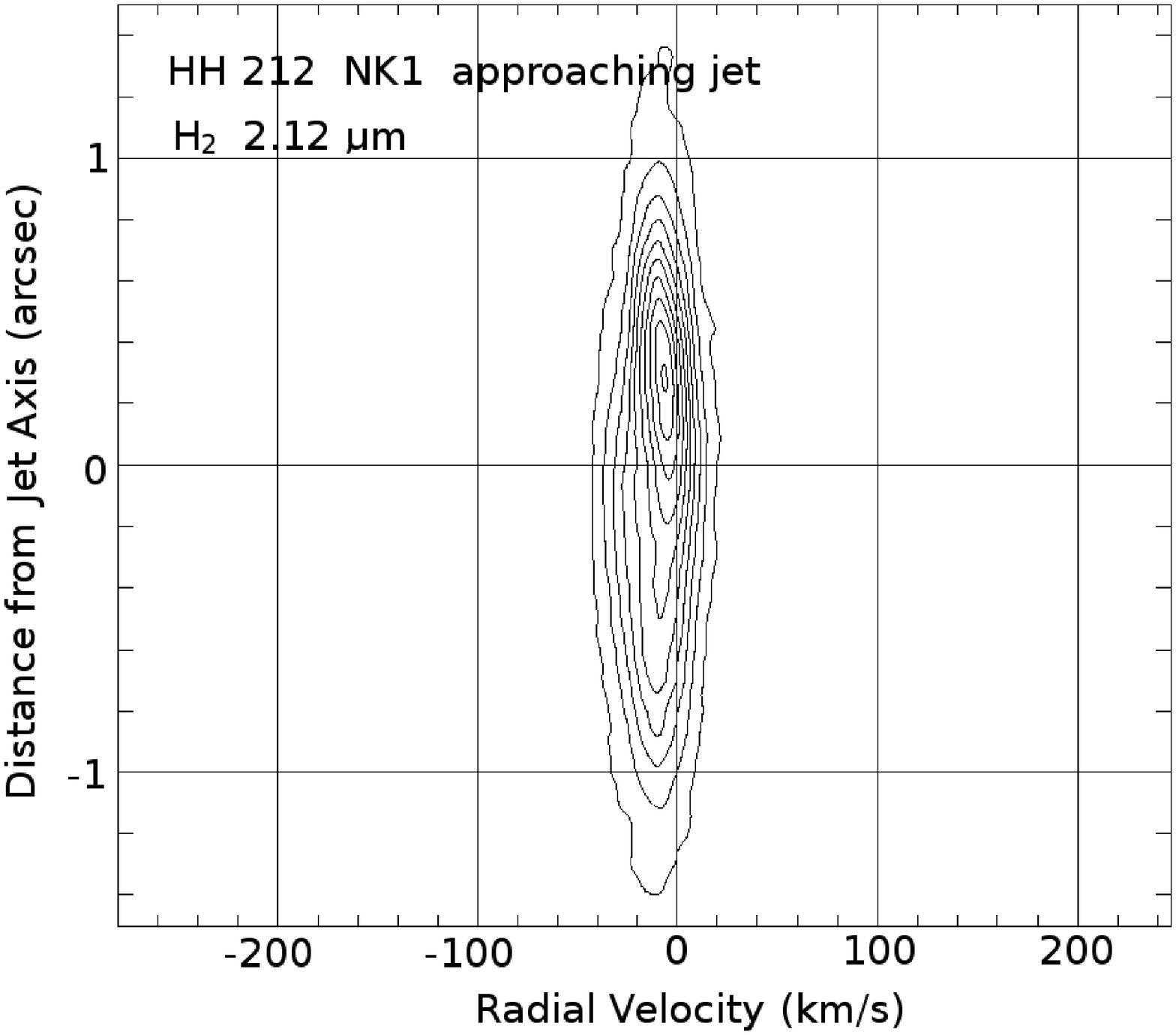}
\includegraphics[width=6.5cm]{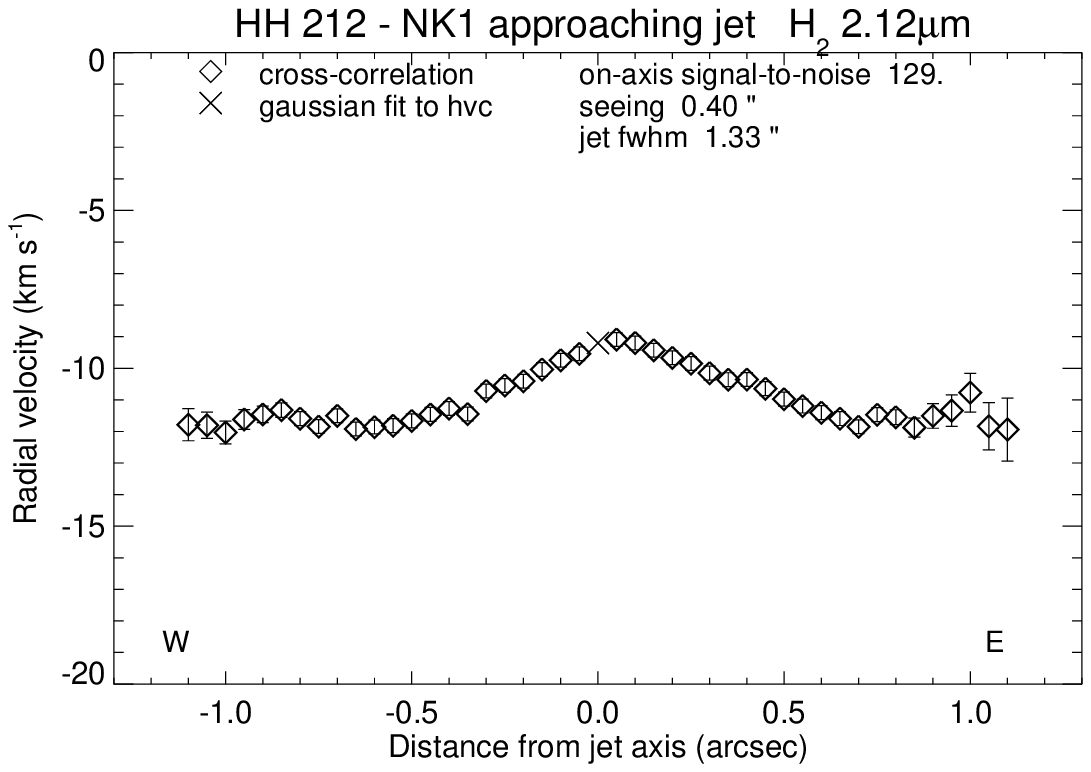}
\includegraphics[width=6.5cm]{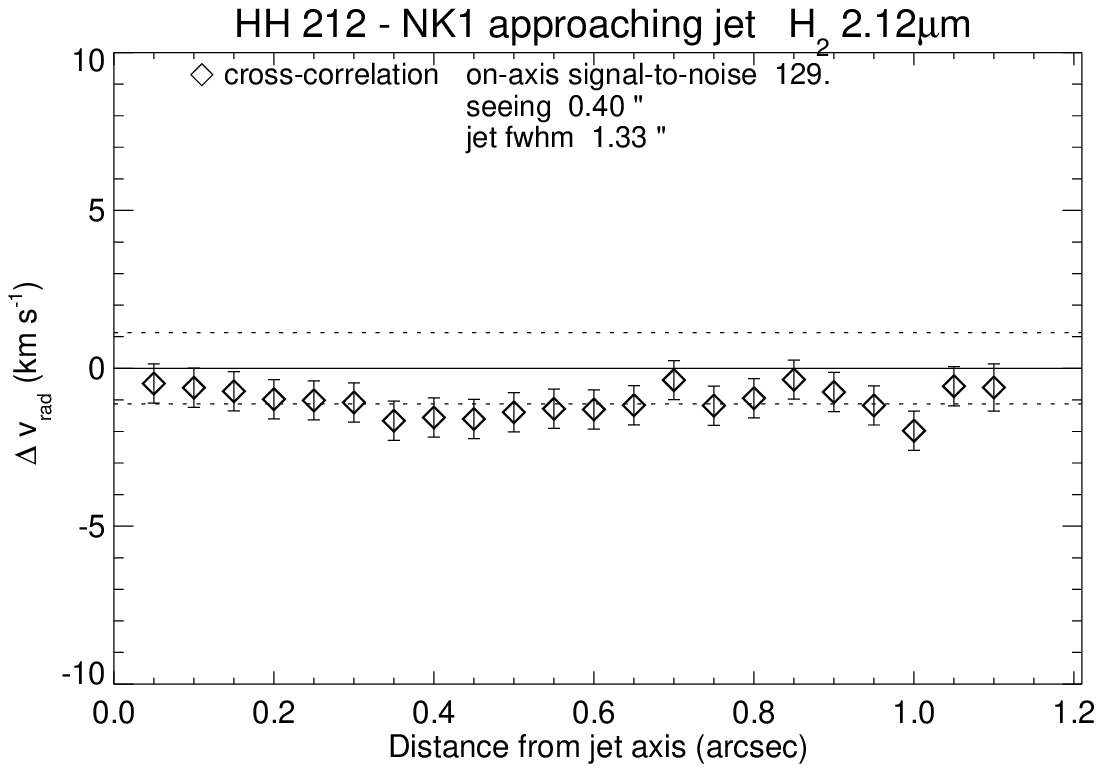}
\caption{Top: position-velocity plots for the approaching lobe of the Class 0 bipolar jet HH\,212, NK1 knot, in [\ion{Fe}{II}] and H$_2$ emission, where positive distances correspond to south-east; Bottom: radial velocity profile across the jet  in the spatially resolved H$_2$ emission, and a plot of the differences in the Doppler shift, $\Delta$v$_{rad}$, between one side of the jet knot and the other. However, caution must be taken in its interpretation given that the emission intensity is asymmetric in the position direction. 
\label{hh212_nk1}}
\end{figure*}

There is a clear detection in both [\ion{Fe}{II}] and H$_2$ emission in the NK1 knot from the HH\,212 outflow of its deeply embedded Class\,0 source, Figure\,\ref{hh212_nk1}.  
[\ion{Fe}{II}] emission is notably fainter with about a third of the signal-to-noise of H$_2$. 
The [\ion{Fe}{II}] HVC is travelling at -27\,($\pm$1)\,km\,s$^{-1}$, 
while the H$_2$ LVC is travelling at -8\,($\pm$1)\,km\,s$^{-1}$, consistant with \citet{Davis00}  and \citet{Correia09}. 
The jet width is unresolved in [\ion{Fe}{II}] emission, but resolved in  H$_2$, and shows a slight hint of a Doppler gradient. Although nearly all $\Delta v_{rad}$ data points are within error bars about zero, there is a systematic trend such that the differences in radial velocity are consistently negative rather than randomly scattered about zero. The Doppler analysis suggests that the south-east side of the flow to be redder than the northwest side. 
However, for NK1, this spatial intensity distribution is asymmetric (as seen in the position-velocity plot), and as also seen in the H$_2$ study by \citet{Correia09}. Recall from Section\,\ref{dataanalysis} that the emission was binned and fitted with a Gaussian in the spatial direction to determine the jet axis, and that this was used as a reference point in determining differences in radial velocity between one side of the jet and the other. Therefore, the spatial asymmetry introduces inaccuracies in determining the location of the jet axis, and hence the velocity differences across the jet. 

\subsection{HH\,212 - SK1}
\label{hh212sk1}

\begin{figure*}
\centering 
\includegraphics[width=6.5cm]{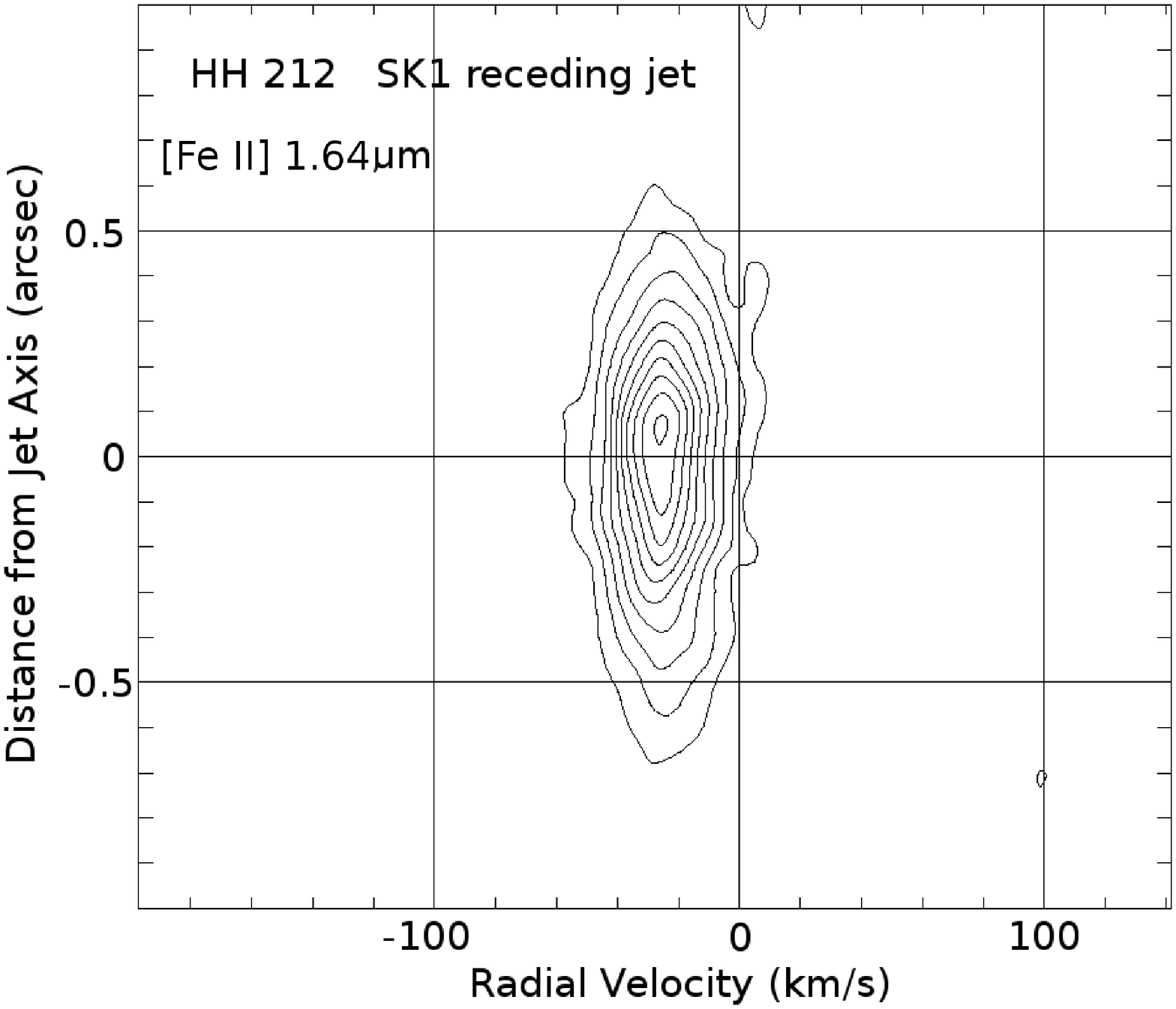}
\includegraphics[width=6.5cm]{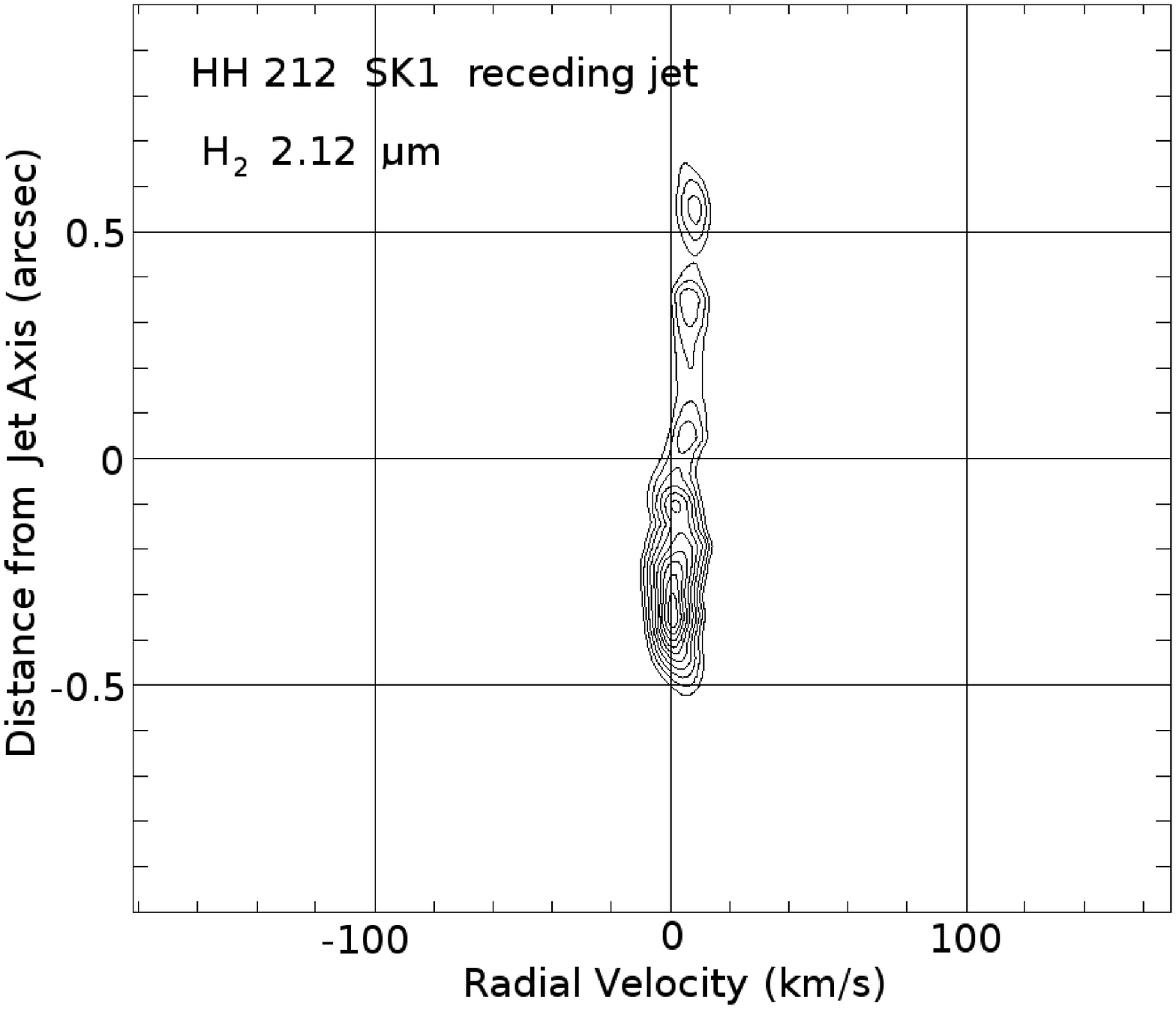}
\includegraphics[width=6.5cm]{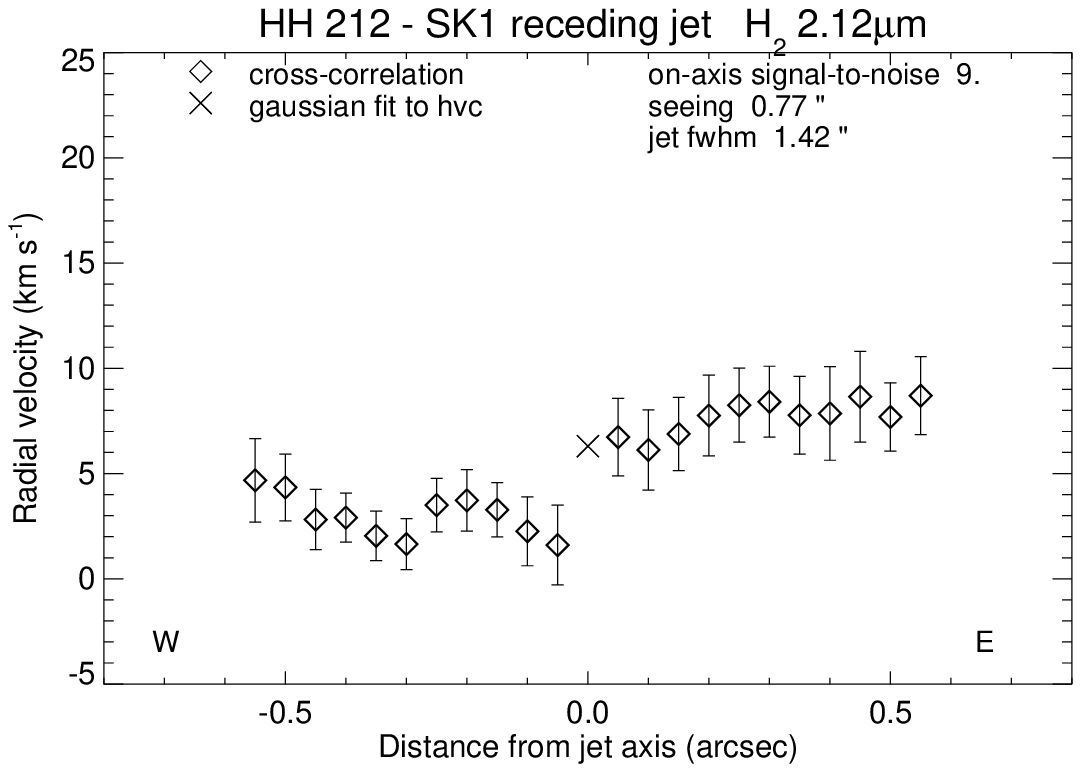}
\includegraphics[width=6.5cm]{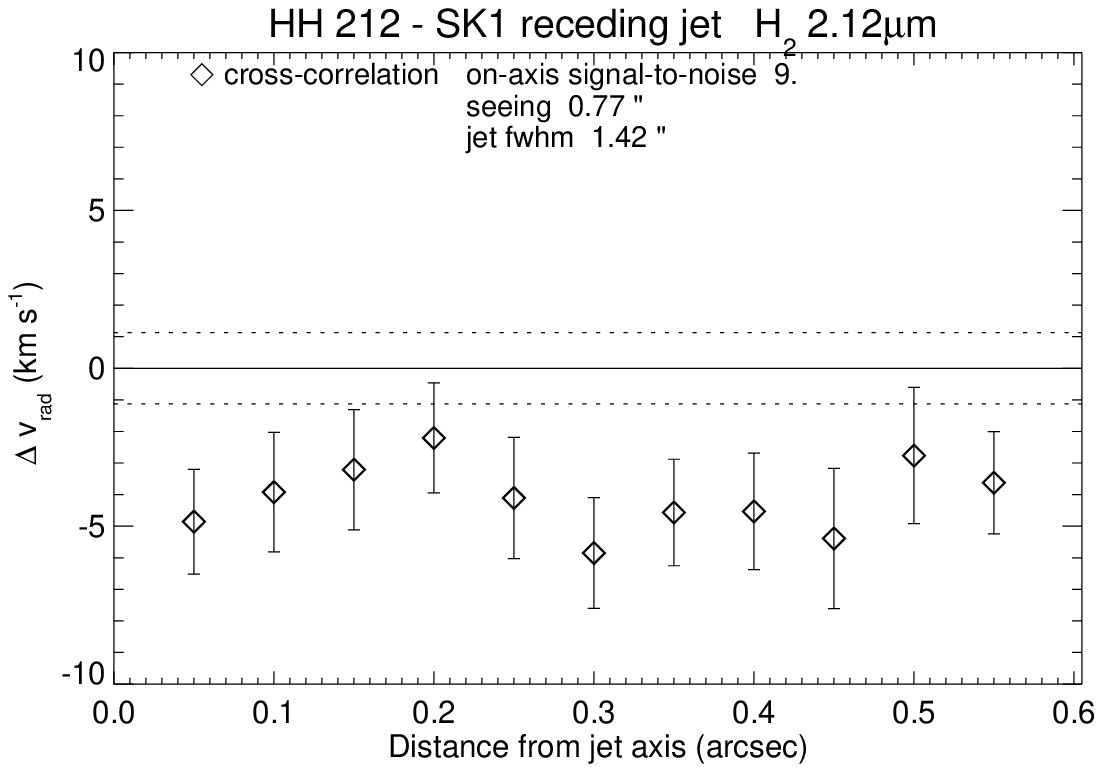}
\caption{Top: position-velocity plots for the approaching lobe of the Class 0 bipolar jet HH\,212, SK1 knot, in [\ion{Fe}{II}] and H$_2$ emission, where positive distances correspond to south-east; Bottom: radial velocity profile across the jet  in the spatially resolved H$_2$ emission, and a plot of the differences in the Doppler shift, $\Delta$v$_{rad}$, between one side of the jet knot and the other. However, caution must be taken in its interpretation given that the emission intensity is asymmetric in the position direction. 
\label{hh212_sk1}}
\end{figure*}

The receding jet knot SK1 from HH\,212 was clearly detected in both [\ion{Fe}{II}] and H$_2$, Figure\,\ref{hh212_sk1}. Again, the jet width in [\ion{Fe}{II}] emission is unresolved. The H$_2$ LVC is travelling at +6\,($\pm$1)\,km\,s$^{-1}$, consistant with \cite{Davis00}. For H$_2$, the values of $\Delta v_{rad}$ are consistently negative, although again we see a spatial asymmetry in the emission (see position-velocity plot). The direction of the Doppler shift reveals the south-east side of the flow to be redder than the northwest side. This matches the hint of a gradient seen in our NK1 dataset (Section\,\ref{hh212nk1}). 
It also matches the direction of the H$_2$ measurements of \citet{Davis00} and \citet{Correia09}. 
and the direction of the Doppler gradient reported for the disk \citep{Wiseman01}. 
However, caution must be taken in the interpretation of the H$_2$ gradient for both NK1 and SK1. 
The emission intensity is asymmetric in the position direction in both lobes, 
but on opposite sides of the jet axis, as also seen in the H$_2$ study by \citet{Correia09}. 

Overall, the results for HH\,212 are in agreement with those obtained by \citet{Davis00} for SK1 with different telescopes and at different epochs (7 years apart). \citet{Davis00} and \citet{Lee08} both report a high positive Doppler slope in SK1 with respect to NK1, and we find the same trend. Therefore, these signatures are real and persist over time. 

Lastly, note that the [\ion{Fe}{II}] HVC is measured as having a radial velocity of -25\,($\pm$1)\,km\,s$^{-1}$. This is highly unusual given that this is the receding jet lobe. To ensure the measurement is correct, we have checked the offsets of the slits in the raw data headers; the LSR velocity correction with respect to the observation dates; the position of the emission line with respect to the OH sky lines in the original raw data images; the wavelength calibration procedure; and the velocity of the H$_2$ line with regard to the literature. No errors were found in the acquisition or calibration of the data, and all other measurements are consistent with previously reported values in the literature. Therefore, this velocity measurement appears to be real but we cannot find an explanation for it based on these data. 

\subsection{Sources of Errors}

Difficulty in determining transverse Doppler profiles arises if the slit is unevenly illuminated, 
e.g. if the target is not well-centered in the slit. 
Using a slit orientated perpendicular to the jet propagation direction (as in our observations) 
ensures that uneven slit illumination is a second order effect, 
and hence the problem of the so-called slit-effect is negligible \citep{Chrysostomou08}. 
This was not the case in previous studies using a slit parallel to the flow 
(i.e. \citealp{Bacciotti02}; \citealp{Woitas05}), 
and so the subtraction of the uneven illumination effect added an element of complexity to the data reduction which we have avoided here. 

\section{Discussion}
\label{discussion}

We have detected gradients in the transverse Doppler-shift profile in resolved H$_2$ jet targets. We now examine whether or not we are justified in interpreting these gradients as a rotation of the flow, and thus whether we can claim to support the magneto-centrifugal mechanism of ejection. We first investigate whether the implied toroidal velocity and angular momentum flux are realistic, in the context of jet and disk rotation measurements reported in the literature. We then examine the statistical foundations for such a claim. 

The implied jet toroidal velocity may be calculated from the measured radial velocity differences. 
The implied toroidal velocity is $v_{\phi}=(\triangle v_{rad}/2)/\cos\,i_{jet}$, where $i_{jet}$ is the inclination angle with respect to the plane of the sky, and $\triangle v_{rad}$ is the radial velocity difference between two sides of the jet. We see a systematic radial velocity difference of $\sim$3\,km\,s$^{-1}$ for the HH\,212 jet. The implied toroidal velocity is then $\sim$1.5\,km\,s$^{-1}$. This is in line with values derived by \citet{Chrysostomou08} of 1 to 7 \,km\,s$^{-1}$ measured for two Class~I sources examined in H$_2$ emission. These authors demonstrated the velocity signature to be consistent with a simple jet rotation model. The value is also in the same range as those measured in H$_2$ emission for HH\,212\,SK1 by \citet{Davis00}, in which the authors point out that the expected jet rotation speed based on the disk rotation measurements of \cite{Wiseman01} is $\sim$2\,km\,s$^{-1}$, thus demonstrating that our measurements are consistent with a jet rotation interpretation. It is also clear from the modeling of \citet{Correia09} that the toroidal velocity may be confused with other kinematic signatures, and so our calculations are broad indications. 

Comparing with more evolved Class~II T\,Tauri jets, our toroidal velocity is substantially lower than those measured in optical and near-UV emission close to the launch point, which yielded typically 10 to 20\,km\,s$^{-1}$ (\citealp{Bacciotti02}; \citealp{Woitas05}; \citealp{Coffey04}; 2007). These results originate from the resolved atomic emission which is more collimated, and so would be expected to yield higher toroidal velocity values with respect to the molecular component. Unfortunately, without resolving the atomic component in our Class~0/I sample, it is not possible to make a direct comparison between the two classes, in order to understand the evolution of angular momentum extraction over the age of the source. 

Nevertheless, we attempt to gain a rough estimate. The angular momentum extraction may be approximated as $\dot L_{jet}$\,$\sim$\,$rv_{\phi}\dot M_{jet}$. For the molecular knot HH\,212 SK1, \citet{Davis00} report $\dot M_{jet}$~$\sim$ 4.2$\times$10$^{-8}$ M$_{\odot}\,$yr$^{-1}$. This is in the range of 2 to 5$\times$10$^{-8}$ M$_{\odot}\,$yr$^{-1}$ reported for Class I outflows \citep{Antoniucci08}. We adopt $v_{\phi}$$=$1.5\,km\,s$^{-1}$ at an average distance from the jet axis of 0$\farcs$4. Although the jet is resolved, we do not see reflected continuum emission which we could use in a PSF deconvolution to determine the true jet radius. These values give $\dot L_{jet}$~$\sim$\,1$\times$10$^{-5}$ M$_{\odot}\,$yr$^{-1}$\,AU\,km\,s$^{-1}$. This is in line with that for HH\,26 of 2$\times$10$^{-5}$ M$_{\odot}\,$yr$^{-1}$\,AU\,km\,s$^{-1}$ derived at 1-2$\arcsec$ from the disk plane, for a jet radius of 0$\farcs$44. However, our result implies a specific angular momentum which is 3 times higher than that of the SiO knot SS \cite{Lee08}, which lies closer to the star at 2$\arcsec$. The difference mainly arises from a smaller adopted SiO jet radius of 0$\farcs$1. 

Roughly comparing Class~0/I values with those for Class~II jets, we take the example of T\,Tauri systems CW\,Tau, RW\,Aur and DG\,Tau, which have been studied close to the jet footpoint, i.e. 0$\farcs$5 (70\,AU) from the disk-plane. \citet{Coffey08} report angular momentum fluxes in one lobe from each system as 1, 3 and 13~$\times$10$^{-6}$ M$_{\odot}\,$yr$^{-1}$\,AU\,km\,s$^{-1}$ respectively for the atomic component. We find up to an order of magnitude difference between the angular momentum extraction of Class~0/I versus Class~II sources. This is a rough indication of the magnitude of the decrease in angular momentum extraction as the young stars evolve. Indeed, the decrease in mass accretion flux supports this:  a comparison between Class~II \citep{Gullbring98} and Class~I \citep{Antoniucci08} shows the mass accretion rates for the Class~I are between one and two orders of magnitudes larger than those of Class~II. 

It seems {\em possible} that we are observing a jet rotation signature in these data. We must also consider whether it is {\em probable}. Does the gradient in the radial velocity profile in the transverse direction represent a rotation of the jet, or could it be a signature of an asymmetric bowshock, precession of the flow, or indeed a combination of effects. 

Steady-state MHD jet models imply that the rotation of the flow should persist to large distances, as a necessary outcome of magnetic flux conservation. Nevertheless, an important observational precaution is to examine the flow close to the launch point. In this way, we try to ensure that environmental factors do not significantly disrupt the inherent jet kinematics. This is not always observationally possible due to high opacity close to the star, especially when dealing with younger more embedded flows as is the case here. Hence the need to observe these younger outflows at the position of the brightest knots, which are not necessarily close to the launch region. This observational obstacle  we must examine the overall context of our results. 

Support for a rotation argument lies in important consistency checks. 
For example, if the jet is rotating as a result of the launch mechanism, we should consistently measure gradients in the jet Doppler profile transverse to the flow direction for many targets. Also, the direction of the implied jet toroidal velocity should match measurements for the direction of the disk rotation within the same system, since the jet is supposedly extracting its angular momentum. Likewise, the implied sense of rotation should be the same in both lobes of a bipolar jet and should persist along the flow. 

From our survey, it appears that transverse Doppler gradients may be consistently measured in Class~0/I jets in the near IR $H_2$ emission, as we set out to establish following the initial findings of \cite{Chrysostomou08}. Furthermore, we provide confirmation of a previous detection for HH\,212 \cite{Davis00}, thus also demonstrating a persistence of the transverse Doppler gradient over time (from 1999 to 2006). Together with the fact that gradients are also consistently measured in Class\,II jets in the optical and near UV, we have an encouraging case so far. 

Next we consider Doppler gradients of jet and disk. Agreement is compulsory to validate a jet rotation interpretation. For HH\,212, we confirm an agreement in sense, as also reported by both \cite{Davis00} for SK1 and \citet{Lee08}. However, they do not agree with the NK1 results of \cite{Davis00} or the SiO results of \citet{Codella07}. 
Furthermore, given the spatial asymmetries we find in this case, our results must be treated with caution. 
We cannot determine a clear gradient in the jet for HH\,111 in order to compare with the disk (\citealp{Yang97}; \citealp{Lee09}), but the data suggest a gradient which opposes that of the disk. Other cases examined are the T\,Tauri systems DG\, Tau (\citealp{Bacciotti02}; \citealp{Testi02}), RW\,Aur (\citealp{Coffey04}; \citealp{Cabrit06}), CW\,Tau (\citealp{Coffey07}) and HH\,30 (\citealp{Coffey07}; \citealp{Pety06}). Agreement has also been confirmed for DG\,Tau, in which jet observations were conducted close to the launch point and where the transverse gradient persists for 100\,AU. Disconcertingly, RW\,Aur shows a disagreement in sense. This may be a complex triple system and hence must be further investigated to confirm that no other influences are coming into play. For CW\,Tau, agreement has been found in preliminary disk results (C. Dougados, private communication). For HH\,30, the jet rotation sense was deemed inconclusive. It has since been revealed that the HH\,30 jet is in fact wiggling \citep{Anglada07}, due to the orbital motion of the binary source \citep{Guilloteau08}. Lastly, agreement has been recently found in both HH\,211 and CB26 (\citealp{Lee07}; \citealp{Launhardt09}). Overall, of the 8 systems for which both jet and disk gradients are studied at this early stage of our work, four show clear agreement and 1 shows clear disagreement. The statistics are as yet too low to be significant. 

Agreement between Doppler gradients in both lobes of a bipolar jet is also compulsory to validate a jet rotation interpretation.  We measure an agreement in the gradient between the two lobes of the HH\,212 bipolar flow. However, we again sound a note of caution based on the spatial asymmetry and the fact that we observe far from the source. This supports the agreement also found in SiO measurements in the two lobes \citep{Lee08}, although the same agreement was not found in \cite{Davis00}. Only two other bipolar flows have been examined. They are from the T\,Tauri systems Th\,28 and RW\,Aur \citep{Coffey04}, and in both cases agreement was found. Furthermore, in two cases were these bipolar jets have been observed with the slit {\em parallel} to the flow, the gradient has persisted {\em along} the jet over a distance of 90 AU in the same direction in both lobes. Such a scenario is not likely to arise from a signature of asymmetric shocking, and certainly not from jet precession where the gradient should be opposite in the two lobes. In other words, in the case of precession, for several slits parallel to the jet (mimiking an IFU), the poloidal velocity peak is offset with respect to the central slit in opposite directions for each lobe, and hence the radial velocity profile shows a gradient opposite in direction in each lobe. In this case, also, there is a periodic change in the direction of the gradient as a function of distance along each jet lobe. Such changes are not found in these parallel slit studies. 

Although statistics are limited, we are building our way towards a statistical argument 
to support the fact that we are indeed observing jet rotation. 
Such information is critical in finally providing observational confirmation  
for the widely accepted but untested centrifugal MHD wind launching mechanism. 
Obviously, any evidence which suggests protostellar jets are 
launched centrifugally would, of course, in turn support the idea that 
the same mechanism is at work in their larger-scale brethren, i.e.\ the 
AGN jets. Indeed, the latest observations indicate jet rotation is also detectable in active galactic nuclei \citep{Young07}. Together with observations indicating jet rotation in protostellar jets, this provides support for the universality of the theory of magneto-centrifugal ejection. 

\section{Conclusions}
\label{conclusions}

We have conducted a ground-based survey in near IR lines 
of four jets from Class 0/I sources 
to search for signatures of jet rotation. 
These embedded sources make the use of adaptive optics very difficult, 
due to the absence of a nearby optical guide star. 
Thus, our results are derived from seeing-limited observations and rely on profile fitting in the spectral direction. 
For HH\,34, we find emission in [\ion{Fe}{II}] and H$_2$ is detected but spatially unresolved with seeing of 0$\farcs$4-0$\farcs$6 across the jet within 1$\arcsec$ of the driving source. For HH\,111-H, we find we cannot resolve a clear gradient across the jet, but perhaps glimpse only a hint of one, which is opposite to the direction of the disk. Similarily, in the NK1 knot of the HH\,212 bipolar flow, we see a hint of a gradient but this time in the same direction as the disk. Lastly, we detect a gradient of typically 2-5\,km\,s$^{-1}$ in SK1 knot of HH\,212. We find agreement in the sense of the gradient in both HH\,212 lobes, as would be expected if the measurements are of jet rotation. The result confirms an earlier detection in SK1 by \citet{Davis00} and \citet{Correia09}, and also matches the disk rotation sense \citet{Wiseman01}. Furthermore, it demonstrates the persistence of the gradient over a time frame of seven years. 

For HH\,212, we can estimate a possible implied toroidal velocities of 1.5\,km\,s$^{-1}$, previously shown to be consistent with a simple jet rotation model \citep{Chrysostomou08}, and angular momentum flux estimates of $\dot L_{jet}$~$\sim$\,1$\times$10$^{-5}$ M$_{\odot}\,$yr$^{-1}$\,AU\,km\,s$^{-1}$. A direct comparison cannot be made between Class~0/I and Class~II angular momentum extraction, since [\ion{Fe}{II}] is unresolved in our data. We are therefore prevented from determining any particular trend over evolutionary time. However, for the HH\,212 and HH\,111 knots, which were observed very far from the driving source, we must consider that the gradients are less likely to be unaffected by external factors such as asymmetric shocking. 

Our analysis illustrates the observational difficulties in conducting this study, as it pushes instrumentation to the limit in its demand for a combination of high spatial and spectral resolution, as well as good sensitivity. Overall, it is clear that in order to safely interpret Doppler gradients as signatures of jet rotation, there is a need for improved statistics via high resolution multi-wavelength jet observations close to the source, and comparison with the associated disk rotation sense. 

\vspace {0.2in}
{\bf Acknowledgements} 
\vspace {0.1in}
\newline
Based on observations at the Gemini Observatory, under Program ID GS-2006B-Q-46, which is operated by the Association of Universities for Research in Astronomy, Inc., under a cooperative agreement
with the NSF on behalf of the Gemini partnership: the National Science Foundation (United
States), the Science and Technology Facilities Council (United Kingdom), the
National Research Council (Canada), CONICYT (Chile), the Australian Research Council
(Australia), CNPq (Brazil) and SECYT (Argentina). 
The present work was supported in part by the European Community's Marie Curie Actions - Human Resource and Mobility within the JETSET (Jet Simulations, Experiments and Theory) network, under contract MRTN-CT-2004-005592. D. C. wishes to acknowledge funding received from the Irish Research Council for Science, Engineering and Technology (IRCSET). 


\end{document}